\documentclass[journal,onecolumn,draftcls,]{IEEEtran}
\ifCLASSINFOpdf
   \usepackage[pdftex]{graphicx}
\else
\fi

\usepackage{soul}
\usepackage{makecell}
\usepackage{subcaption}
\usepackage{xcolor}
\usepackage{nomencl}
\usepackage{mathtools, cuted}
\usepackage{mathtools}
\usepackage[normalem]{ulem}
\makenomenclature
\usepackage{multicol}
\usepackage{etoolbox}
\usepackage{amsmath,esint}
\renewcommand\nomgroup[1]{%
  \item[\bfseries
  \ifstrequal{#1}{A}{Symbols}{%
  \ifstrequal{#1}{B}{Subscripts}{%
  \ifstrequal{#1}{C}{Superscripts}{}}}%
]}

\makeatletter
\newcommand*{\rom}[1]{\expandafter\@slowromancap\romannumeral #1@}
\makeatother

\begin{document}
%
\title{Validity of solid-state Li$^+$ diffusion coefficient estimation by electrochemical approaches for lithium-ion batteries}












\author{Zeyang Geng$^1$, Yu-Chuan Chien$^2$, Matthew J. Lacey$^3$, Torbj{\"o}rn Thiringer$^1$, and Daniel Brandell$^2$

\thanks{$^1$ Department of Electrical Engineering, Division of Electric Power Engineering, Chalmers University of Technology, SE-412 96 Göteborg, Sweden (e-mail: zeyang.geng@chalmers.se)

$^2$ Department of Chemistry - Ångström Laboratory, Uppsala University, SE-751 21 Uppsala, Sweden

$^3$ Scania CV AB, SE-15187 Södertälje, Sweden}
}

\IEEEpubid{}


\maketitle

\begin{abstract}

The solid-state diffusion coefficient of the electrode active material is one of the key parameters in lithium-ion battery modelling. Conventionally, this diffusion coefficient is estimated through the galvanostatic intermittent titration technique (GITT). In this work, the validity of GITT and a faster alternative technique, intermittent current interruption (ICI), are investigated regarding their effectiveness through a black-box testing approach. A Doyle-Fuller-Newman model with parameters for a LiNi$_{0.8}$Mn$_{0.1}$Co$_{0.1}$O$_2$  electrode is used as a fairly faithful representation as a real battery system, and the GITT and ICI experiments are simulated to extract the diffusion coefficient. With the parameters used in this work, the results show that both the GITT and ICI methods can identify the solid-state diffusion coefficient very well compared to the value used as input into the simulation model. The ICI method allows more frequent measurement but the experiment time is 85 \% less than what takes to perform a GITT test. Different fitting approaches and fitting length affected the estimation accuracy, however not significantly. Moreover, a thinner electrode, a higher C-rate and a greater electrolyte diffusion coefficient will lead to an estimation of a higher solid-state diffusion coefficient, generally closer to the target value.










\end{abstract}


%
\IEEEpeerreviewmaketitle

\section{Introduction}


A lithium (Li) ion battery is a complicated electrochemical system and its performance is dependent on a multitude of material properties, among which the solid-state diffusion coefficient $D_s$ of Li$^+$ is one of the key parameters, since the mass transport in these particles is the rate-limiting processes for thin electrodes, and the corresponding resistances constitute a major part of the total battery overpotential. The value of $D_s$ is, however, not straight-forwardly obtained, which is a major obstacle for more precise battery modelling. While the diffusion coefficient can be theoretically estimated from density functional theory (DFT) \cite{zheng2011diffusion,dathar2011calculations} or Molecular Dynamics (MD) simulations \cite{islam2000ionic,islam2014lithium}, it is more frequently experimentally determined by electrochemical approaches, including the galvanostatic intermittent titration technique (GITT), potentiostatic intermittent titration technique (PITT), electrochemical impedance spectroscopy (EIS) and cyclic voltammetry (CV) \cite{santos2018revisiting}. Of these, GITT is arguably the most popular choice due to its comparative robustness, while still being time-consuming. 

GITT was original proposed by Weppner and Huggins \cite{weppner1977determination} for a Li$_3$Sb electrode, and then later generalized for porous insertion electrode materials used in lithium-ion batteries \cite{chen2020development,nickol2020gitt,dees2009analysis}. In the GITT test, a current pulse is applied when the cell is in an equilibrium state and the voltage response follows a linear relationship with the square root of time. The diffusion coefficient can thus be calculated from the voltage response, assuming that the particle size is known. Normally the current pulses are applied at different states of charge (SOC) as the solid-state diffusion coefficient is a SOC dependent parameter. Since an equilibrium initial condition is required before the current pulse, a long relaxation time is needed in between each SOC level, which makes the GITT method very time consuming. In recent years, however, the alternative intermittent current interruption (ICI) technique has been proposed \cite{lacey2017influence} as a faster alternative to GITT \cite{chien2021fast}. Using ICI, the voltage response during a current interruption is monitored, which also follows a linear relationship with the square root of time. Since the ICI method does not require the cell to be in an equilibrium state, the most time consuming part in the GITT method, i.e. the relaxation period, can thus be omitted.

Employing the same experimental object and setup, different methods can sometimes provide fairly similar results for the diffusion coefficient $D_s$. For example, Chien et al. obtained diffusion coefficient values with the GITT and ICI methods with an excellent agreement with each other \cite{chien2021fast}, and a similar comparison has been made between the GITT and EIS methods displaying a good match \cite{deng2020consistent}. Nevertheless, the reported $D_s$ values from different literature of one specific material often vary within several orders of magnitude and this fact makes it challenging to decide a proper value to use in battery modelling. The reason for this large spread of the reported $D_s$ values can be explained by the differences in the electrode design, electrolyte composition, current magnitude, fitting method, resting period, etc. Moreover, the validity of GITT, as well as other techniques employed for this purpose, is being questioned and discussed as the underlying assumptions are not completely valid when they are applied for porous insertion lithium-ion electrodes \cite{liu2006analytical,garcia2018quantitative}. Dees et al. has shown that the $D_s$ value estimated with the GITT theory is generally lower that the value used in their DC electrochemical model \cite{dees2009analysis}.

It is not feasible to strictly validate the electrochemical approaches for estimating the solid-state Li$^+$ diffusion coefficient experimentally, since it is difficult to have a diffusion coefficient value that can be trusted as the truth. However, if these electrochemical approaches are valid for real batteries, they shall as well be valid for continuum models of Li-ion batteries, as a model is a simplification of the real world but at the same time keeps all the assumptions used in the electroanalytical methods. The philosophy in this work is that if experiments with these electrochemical methods are simulated with a valid battery model, one should be able to extract the same solid-state diffusion coefficient as what is put in the model, as a proof of the method validity. This is essential so that these electrochemical methods can be trusted to parameterize accurate models.

Therefore, a theoretical validation is demonstrated in this work for the GITT and ICI method when applying them for a porous electrode. The classical Doyle-Fuller-Newman (DFN) model \cite{doyle1993modeling,fuller1994simulation} is one of the most commonly employed continuum models for simulating lithium-ion batteries and it is considered to be a fairly faithful representation of a real battery system. In this work, a DFN model previously implemented in MATLAB \cite{geng2021bridging} with parameters from \cite{chen2020development} for LiNi$_{0.8}$Mn$_{0.1}$Co$_{0.1}$O$_2$ is used for black-box testing. The input is current sequences designed according to the GITT and ICI methods and the output is the battery terminal voltage which is used to estimate the solid-state diffusion coefficient. The advantage is that after the black-box testing, different physical phenomenon in the model can be studied in details to demonstrate the origins for any deviations between the estimated $D_s$ value and the true value used in the model. The main purpose of the study is thus to reveal how the porous electrode structure will affect the solid-state diffusion coefficient estimation when employing the GITT and ICI methods, which was originally proposed for a more uniform electrode material. This work also demonstrates the impact on the $D_s$ estimation accuracy with different fitting methods, fitting lengths, current magnitude, electrode thickness as well as electrolyte properties.









\section{Theory}

Following the DFN model, the mass transport process in the electrode particles is modelled by Fick's law in a spherical coordinate system
\begin{equation}
\frac{\partial c_s}{\partial t} = D_s (\frac{\partial ^2 c_s}{\partial r^2} + \frac{2}{r} \frac{\partial c_s}{\partial r})
       \label{eq:Ficks_law}
\end{equation}
where $c_s$ is the lithium-ion concentration in the solid active material, $t$ is the time, $D_s$ is the solid-state diffusion coefficient of lithium ions, and $r$ is the distance from the particle center. When the cell is in an equilibrium state
\begin{equation}
    c_s|_{t=0} = c_0
\end{equation}
If a constant current $j$ is applied to the surface of a particle at $t$ = 0 s, then the lithium ion flux at the surface is
\begin{equation}
 \frac{\partial c_s}{\partial r}|_{r = r_s} = -j/(FD_s)
\end{equation}  
where $F$ is the Faraday constant. The symmetric shape of the particle fulfills the condition
\begin{equation}
 \frac{\partial c_s}{\partial r}|_{r=0} = 0
\end{equation}
The analytical solution of (1) with the boundary conditions in (2)-(4) is
\begin{equation}
    c_{s,surf} = c_0 + \frac{j r_s}{F D_s} f(\frac{D_s t}{r_s^2})
\end{equation}
\begin{equation}
    f(\frac{D_s t}{r_s^2}) = f(\tau) = 3 \tau +0.2 -2 \sum_{n=1}^{\infty} \frac{1}{\lambda _n^2} exp(-\lambda _n^2 \tau)
\end{equation}
where $\lambda _n$ is the positive roots of $\lambda  = \tan (\lambda)$ \cite{nickol2020gitt}. The expression in (6) can be approximated to be
\begin{equation}
  f(\tau) =
    \begin{cases}
      \frac{2}{\sqrt{\pi}} \sqrt{\tau} & \tau \ll 1\\
      3 \tau +0.2 & \tau \gg  1
    \end{cases}       
\end{equation}

\begin{figure}[!ht]
    \centering
    \includegraphics{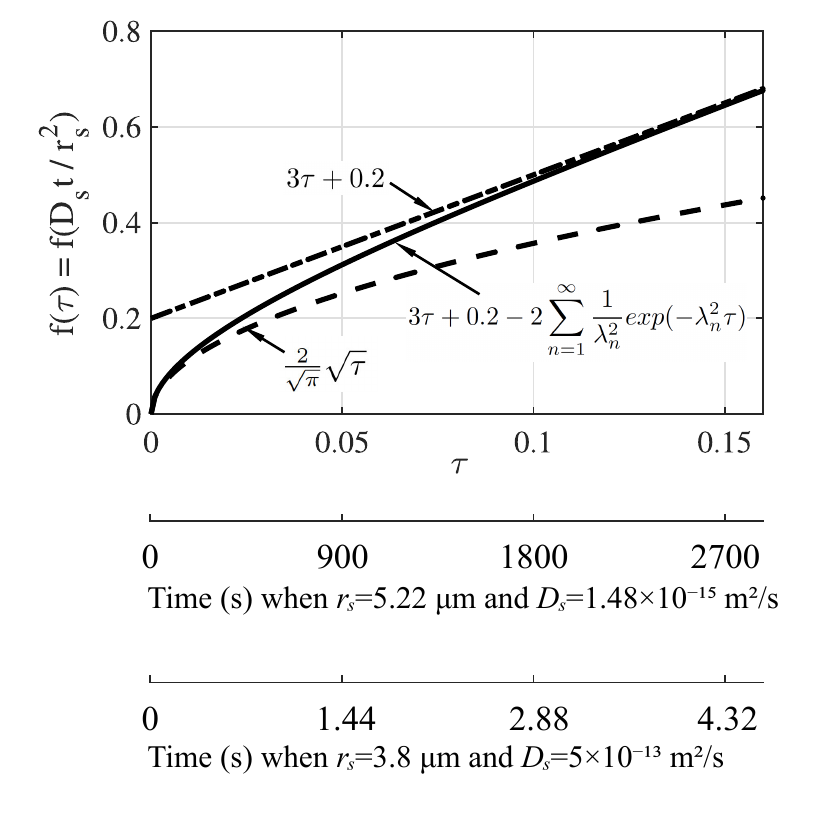}
    \caption{The function in (6) is plotted as a solid line (n = 1272 for $\lambda _n$), while its approximations in (7) are plotted as dashed lines.}
    \label{fig:Fickslaw_solution}
\end{figure}

According to the mathematical description above, the diffusion coefficient $D_s$ could be found from the step response of $c_s$. In most cases, the concentration is very difficult to measure directly, therefore the terminal voltage $E$ - the experimentally easily measurable parameter - is used instead. The step response (5) is therefore rewritten as
\begin{equation}
    E = E_0 + \frac{dE}{dc} \frac{j r_s}{F D_s} f(\frac{D_s t}{r_s^2})
\end{equation}
In (8), $dE/dc$ is often assumed to be a constant within a small SOC interval, i.e. the voltage profile is linear
\begin{equation}
    \frac{dE}{dc} = \frac{\Delta E}{\Delta c} = \frac{\Delta E_{GITT}}{It_{pulse} /(FL\epsilon_s)}
    \label{eq:dEdc}
\end{equation}
where $I$ is the current density on the electrode, $t_{pulse}$ is the time length of the current pulse, $L$ is the electrode length, $\epsilon_s$ is the volume fraction of the solid, and $\Delta E_{GITT}$ is the voltage difference between two equilibrium states before and after the current pulse. Furthermore, the local current density on particle $j$ in (8) is $j = I/(S_a L)$, where $S_a = 3 \epsilon_s/r_s$ is the specific surface area. Combining (8) and (9) gives
\begin{equation}
    E = E_0 + \frac{ \Delta E_{GITT}}{t_{pulse}} \frac{r_s^2}{3D_s} f(\frac{D_s t}{r_s^2})
    \label{eq:full_expression_fitting}
\end{equation}
 When $\tau = D_st/r_s^2 \ll 1$, \eqref{eq:full_expression_fitting} can be approximated as
\begin{equation}
    E = E_0 + \frac{2}{3}\frac{\Delta E_{GITT}}{t_{pulse}} \frac{r_s \sqrt{t}}{\sqrt{\pi D_s}}
\end{equation}
\noindent and thus the diffusion constant can be found to be
\begin{equation}
    D_s = \frac{4}{9 \pi} (\frac{r_s}{dE/d\sqrt{t}} \frac{\Delta E_{GITT}}{t_{pulse}})^2
    \label{eq:tau<1 fitting}
\end{equation}
The approach in \eqref{eq:tau<1 fitting} is one of the most commonly used fitting methods employed for GITT, and typically only a chosen region of the voltage step response is used to fit $dE/d\sqrt{t}$ to fulfill the approximation condition $\tau \ll 1$. The validity of this approximation is demonstrated in Fig.~\ref{fig:Fickslaw_solution}. It can be observed that the system response follows a linear relationship with $\sqrt{\tau}$ in the beginning and this approximation is considered to be valid when $\tau <$ 0.0032 \cite{chien2021fast}. The time axis is shown for two sets of parameters reported for LiNi$_{0.8}$Mn$_{0.1}$Co$_{0.1}$O$_2$: $r_s$ = 5.22 $\mu$m, $D_s = 1.48 \times 10^{-15}$ m$^2$/s \cite{chen2020development} and $r_s$ = 3.8 $\mu$m, $D_s = 5 \times 10^{-13}$m$^2$/s \cite{sturm2019modeling}. In the first case, $\tau <$ 0.0032 corresponds to $t = \tau r_s^2/D_s<$ 58.92 s and therefore the voltage response within this time length can be used to fit $dE/d\sqrt{t}$ in \eqref{eq:tau<1 fitting} for the diffusion coefficient identification. In contrast, the condition $\tau <$ 0.0032 corresponds to $t <$ 0.09 s in the latter case, meaning that the $\tau \ll 1$ approximation fails to describe the system response almost instantly. For such material properties, the expression in \eqref{eq:full_expression_fitting} can be used instead, which is not limited by the time constant of the system \cite{delacourt2011measurement}. In this work, both of the two methods will be examined. Moreover, it should be acknowledged that there exist other less common variants to estimate the $\Delta E/\Delta c$ in (8), resulting in similar expressions as those described above \cite{nickol2020gitt,wei2015kinetics,shen2013least}, but those variants will not be further discussed here.

\begin{figure}
    \centering
    \includegraphics{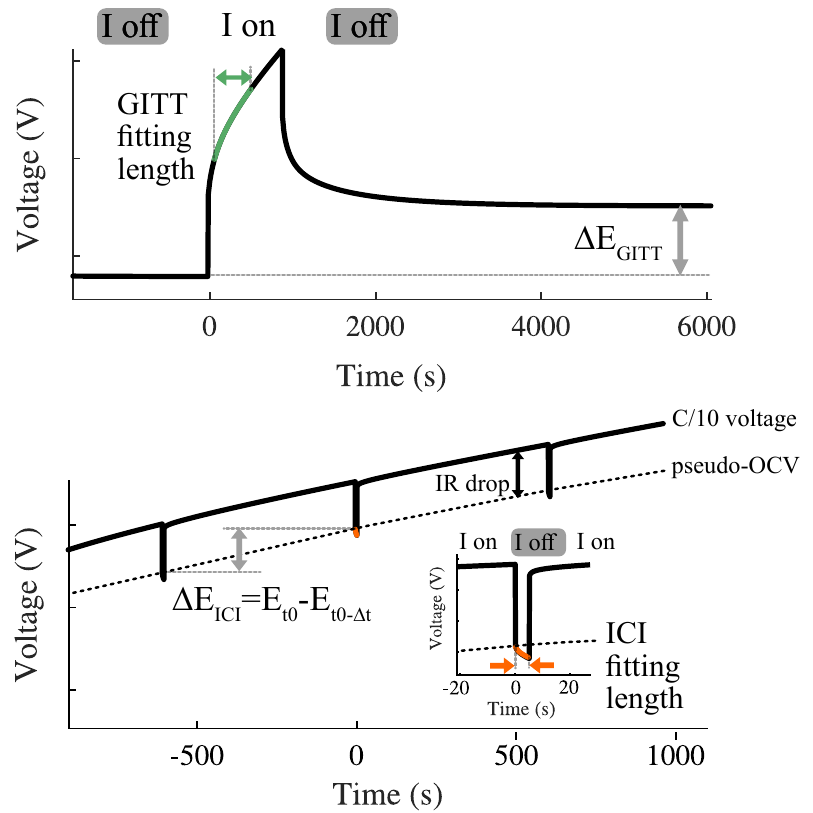}
    \caption{The voltage response during one current pulse, comparing the GITT (above) and ICI (below) approaches.}
    \label{fig:voltage_profile_illustration}
\end{figure}

In the GITT method, the solution of voltage response in (8) is solved with the initial condition where the cell is in an equilibrium state, meaning that a sufficient relaxation time period needs to take place between the current pulses. This generally very long time consumption has been considered to be the main disadvantage of GITT when compared with other methods. Alternatively, the more efficient ICI technique \cite{lacey2017influence} can be considered as a possible replacement. In ICI, the step response when the current is switched off is utilized. When the current is stopped at $t = t_1$, the concentration on the surface is

\begin{equation}
    c_{s,surf} = c_0 + \frac{j r_s}{F D_s} f(\frac{D_s t}{r_s^2})-H(t)\frac{j r_s}{F D_s} f(\frac{D_s (t-t_1)}{r_s^2})
\end{equation}

\begin{equation}
  H (t) =
    \begin{cases}
      0 & t<t_1\\
      1 & t\geq t_1
    \end{cases}       
\end{equation}
where $t$ is counted from the time when the current is switched on \cite{chien2021fast}. Similar as in the GITT method, the initial concentration response when the current stops also follows a linear relationship with $\sqrt{t}$ according to the approximation in (7). In principle, the diffusion coefficient $D_s$ could be derived in a similar manner as in \eqref{eq:tau<1 fitting}, resulting in the expression 
\begin{equation}
    D_s = \frac{4}{9 \pi} (\frac{r_s}{dE/d\sqrt{t}} \frac{\Delta E_{ICI}}{\Delta t})^2
    \label{eq:ICI_fitting}
\end{equation}
where $dE/d \sqrt{t}$ is fitted during the zero current period and $\Delta E_{ICI}/ \Delta t$ is estimated from a pseudo open circuit voltage, which is obtained by subtracting the IR drop from the voltage curve, as shown in Fig.~\ref{fig:voltage_profile_illustration}. Moreover, the ICI method does not require an equilibrium initial condition as in the GITT method, and the cell can instead be in a dynamic state under a constant current before the current interruption. This feature renders the ICI method considerably less time consuming.

It should be acknowledged that the derivation above is based the following assumptions:
\begin{itemize}
    \item Fick's law is applied.
    \item The particle has a spherical shape and the diffusion is symmetric.
    \item The voltage profile is linear during a short interval, i.e. $dE/dc$ is a constant.
    \item The current density at the particle surface is constant.
    \item The over-potential caused by other electrochemical processes can be ignored.

\end{itemize}
The first two assumptions are the same as in the general DFN model, and their impact is thus not examined. The impact of the latter three assumptions will be studied hereafter to verify the validity of the GITT and ICI methods when they are applied for a porous lithium ion electrode.

\section{Simulation setup}

The mathematical model is a DFN model implemented in MATLAB as described in \cite{geng2021bridging} and the parameters are from \cite{chen2020development}, listed in Table \ref{tab:parametersNMC811}. Two values are here adjusted from the original values reported in \cite{chen2020development}; one is the electronic conductivity of the electrode material and the other is the reaction rate constant in the charge transfer reaction. The electronic conductivity of LiNi$_{0.8}$Mn$_{0.1}$Co$_{0.1}$O$_2$ (NMC811) is reported to be 0.18 S/m in \cite{chen2020development}. However, conductive carbon additives are often added in cell manufacturing to improve the conductivity of the electrode. Therefore, a higher value 10 S/m is used in this work to be closer to reality. The other parameter, the reaction rate constant in the charge transfer reaction, is reduced to increase the charge transfer resistance and thus improve the stability of the system. Instability issues appear under a certain parameter combination with solvers used in \cite{geng2021bridging} but can be overcome by slightly adjusting the parameters. This adjustment is considered to not affect the conclusions in this study for the solid-state diffusion coefficient identification.

A symmetric cell setup is employed: NMC811 ($x_0=x_{min}$) - separator - NMC811 ($x_0=x_{max}$). When the cell is charged, one electrode NMC811 ($x_0=x_{min}$) is being lithiated while the other electrode NMC811 ($x_0=x_{max}$) is being delithiated. This setup has the advantage that the lithiation and delithiation processes of one electrode material can be simulated within one simulation, thereby saving computational time. The electrolyte potential in the middle of the separator is used as the reference potential (ground) to mimic a three-electrode cell. The DFN model is a pseudo two-dimensional (P2D) model where one dimension is in the cross-section direction of the cell to address the current distribution in the electrodes, and the electrolyte concentration distribution through the electrodes and the separator, while another dimension is in the radius direction of the particles for the solid-state diffusion processes. Both dimensions are evenly meshed, with 10 mesh elements in each electrode and the separator and 100 mesh elements in each particle. The time step is 0.1 s during the region used for fitting, and 5 s for the rest to shorten the simulation time.

\begin{table}[!ht]
    \centering
    \caption{Parameters used in the simulation \cite{chen2020development}. $^*$\textit{Adjusted}}
    \begin{tabular}{| c | c |} \hline 
         Parameters &  Values \\ \hline
         L   &   75.6 $\mu $m\\
         $r_s$  &    5.22 $\mu $m\\
         $\epsilon_l$ &    0.335\\
         $\epsilon_s$ &    0.665\\
         Bruggeman constant  & 2.43\\
         \hline
         Open circuit voltage (OCV) & \eqref{eq:ocv}\\
         $\sigma_s$  & 10 S/m (original value 0.18 S/m)$^*$  \\
         $D_s$  & 1.48$\times$10$^{-15}$ m$^2$/s \\
         $c_{s,max}$   & 51.77 mol/dm$^3$\\
         $x_{min}$ - $x_{max}$ &  0.266-0.908\\
         $k_a = k_c$  & 7$\times$10$^{-12}$ (original value 3.54$\times$10$^{-11}$ )$^*$\\
         $\alpha_a = \alpha_c$  &  0.5\\
         \hline
         $c_{l,0}$   &  1 mol/dm$^3$\\
         $\epsilon_l$   & 0.47\\
         $D_l$  &  1.77$\times$ 10$^{-10}$ m$^2$/s\\
         $\kappa$  &   0.95 S/m  \\
         $t_+^0$  &  0.26\\
         $f_A$  &  0 \\
         \hline
    \end{tabular}
    \label{tab:parametersNMC811}
\end{table}

\begin{equation}
\begin{split}
OCV(x) = & -0.8090x + 4.4875 - 0.0428 \times \tanh(18.5138(x-0.5542)) - \\
 & 17.7326\times \tanh(15.789(x-0.3117))+ 17.5842\times \tanh(15.9308(x-0.312))
\end{split}
\label{eq:ocv}
\end{equation}



The applied current is 0.1 C = 0.498 mA/cm$^2$ for both the GITT and ICI simulations. In the GITT simulation, the current pulse is 900 s with 1.5 h of relaxation in between. In reality, lithium-ion cells take longer time to reach equilibrium again after the pulse, but 1.5 h is sufficient in these simulations. For the ICI procedure, the current is interrupted for 5 s every 5 min.

Three identification approaches are used to identify the diffusion coefficient $D_s$

\begin{itemize}
  \item GITT full expression fitting according to \eqref{eq:full_expression_fitting}
  \item GITT $\tau \ll 1$ approximation fitting according to \eqref{eq:tau<1 fitting}
  \item ICI fitting according to \eqref{eq:ICI_fitting}
\end{itemize}

In \eqref{eq:tau<1 fitting} and \eqref{eq:ICI_fitting}, $dE/d \sqrt(t)$ is fitted within a selected region, as shown in Fig.\ref{fig:voltage_profile_illustration}. The starting point of the fitting region is one second after the current is switched in order to avoid the transient period of other electrochemical processes. In reality, this can be adjusted to a later point in time to exclude the effect of double layer capacitance. Different fitting lengths are used in the data processing. With the full expression fitting in \eqref{eq:full_expression_fitting}, the fitting length is unlimited. However, with the $\tau = D_st/r_s^2 \ll 1$ fitting in \eqref{eq:tau<1 fitting} and \eqref{eq:ICI_fitting}, the fitting length should not be too long to diverge from the conditions of approximation.

\section{Results and discussions}

\begin{figure}[!ht]
    \centering
    \includegraphics{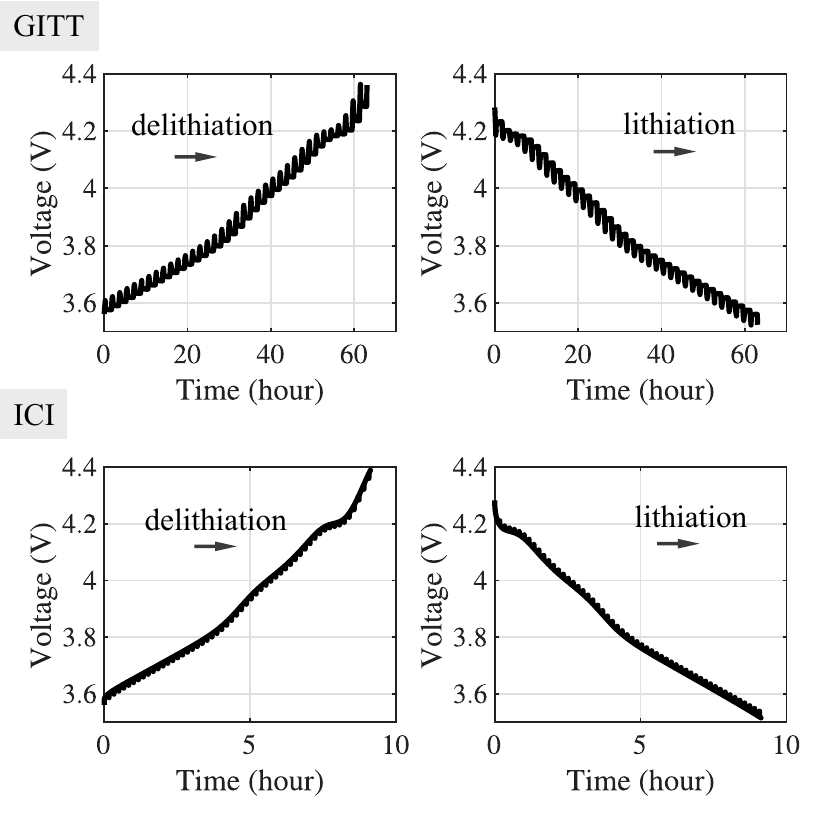}
    \caption{Simulated voltage profiles of an NMC811 electrode during lithiation and delithiation processes with GITT and ICI methods.}
    \label{fig:GITT_ICI_full_soc_voltage}
\end{figure}

The simulated voltage profiles during the GITT and ICI tests are shown in Fig.~\ref{fig:GITT_ICI_full_soc_voltage}, which resemble the experimental results reported in literature \cite{chien2021fast,dees2009analysis}. With the symmetric cell setup described in the previous section, the delithiation and lithiation processes are simulated simultaneously. In the GITT test, the voltage step response is clearly visible under a C/10 current pulse and in total 35 current pulses are applied through the entire SOC range. With the simulation protocol, i.e. 15 minutes pulse with 1.5 hours relaxation, the time required to complete a GITT test in a real experiment will be 63 hours. On the other hand, 53 current interruptions take place in the ICI test, with the corresponding experiment time being merely 9 hours. The ICI method thereby allows more identification occasions for the diffusion coefficient, with less than 15 \% of the time it takes to complete a GITT measurement. 

\begin{figure}
     \centering
     \begin{subfigure}[b]{0.49\textwidth}
    \centering
    \includegraphics{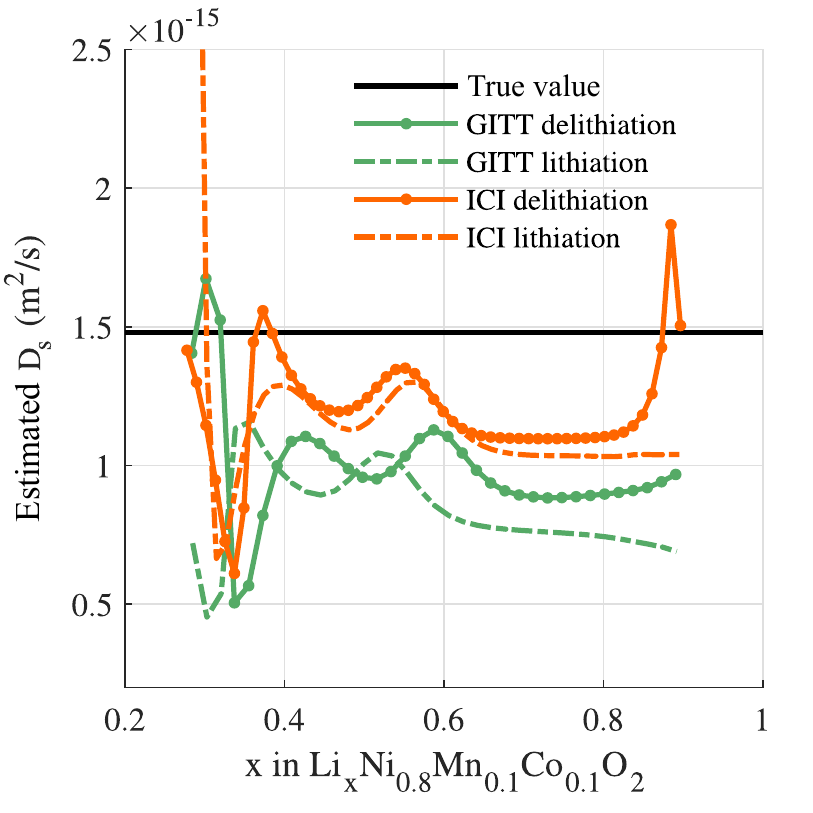}
     \caption{}
     \label{fig:GITT_ICI_full_soc_D_s}
     \end{subfigure}
     \hfill
     \begin{subfigure}[b]{0.49\textwidth}
    \centering
    \includegraphics{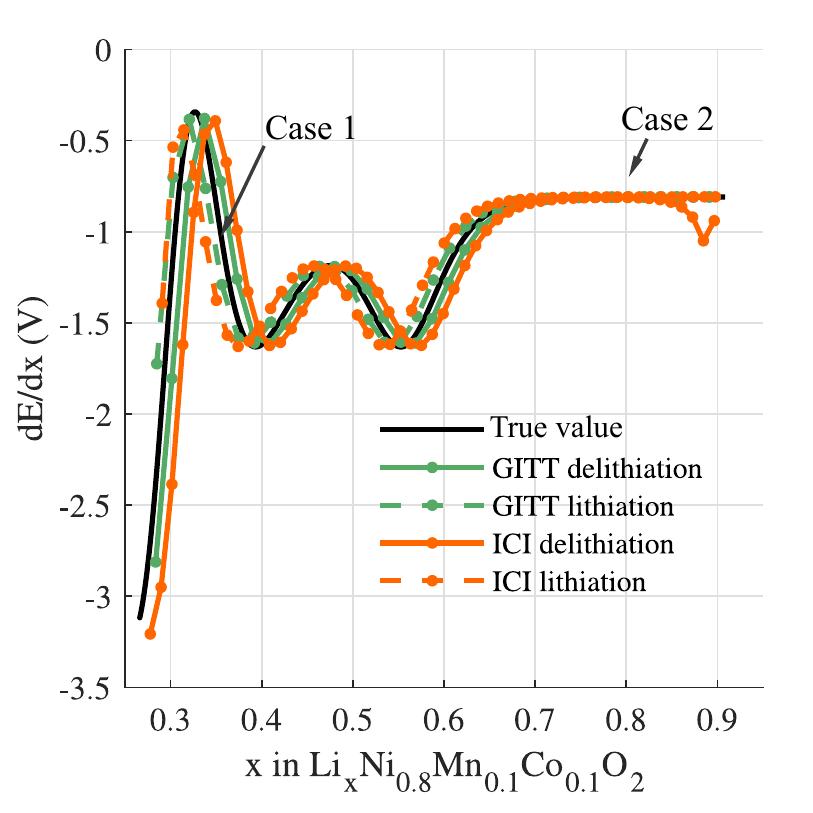}
     \caption{}
    \label{fig:GITT_ICI_full_soc_dEdx}
     \end{subfigure}
    \caption{(a) The derived diffusion coefficient $D_s$ from the GITT and ICI simulations. (b) Estimated $dE/dx$ from the GITT and ICI simulations.}
\end{figure}

The derived diffusion coefficient $D_s$ is presented in Fig.~\ref{fig:GITT_ICI_full_soc_D_s}. In the GITT method, the voltage response between $t$ = 1 s and $t$ = 21 s (the current is applied at $t$ = 0 s) is used to for the fitting in \eqref{eq:tau<1 fitting}, and the voltage response between $t$ = 1 s and $t$ = 5 s (the current is interrupted at $t$ = 0 s) is used in the ICI method to fit with the expression in \eqref{eq:ICI_fitting}. The diffusion coefficient used in the model is 1.48$\times$10$^{-15}$ m$^2$/s \cite{chen2020development} and it is plotted as the 'true value' in Fig.~\ref{fig:GITT_ICI_full_soc_D_s}. Although in reality the diffusion coefficient of lithium ions is SOC dependent for most electrode materials \cite{persson2010lithium}, the parameter value used in this work is a constant for simplicity and for making a clear demonstration. 
It can be seen that the two methods can estimate the diffusion coefficient $D_s$ successfully, with only small deviations from the value used in the model. Since the true value of $D_s$ in the model, as well as other electrochemical parameters (see Table 1), are constant, ideally the estimated $D_s$ value from the GITT and ICI methods should be a constant as well. However, as it is shown in Fig.~\ref{fig:GITT_ICI_full_soc_D_s}, the estimated $D_s$ from the GITT and ICI methods has a significant dependence on the SOC. It can be observed that the trend of the estimated $D_s$ in Fig.~\ref{fig:GITT_ICI_full_soc_D_s} has a good correlation with the $dE/dx$ profile in Fig.~\ref{fig:GITT_ICI_full_soc_dEdx}, and this is a hint that the deviations are related with the violation of one assumption in the mathematical derivation: \textit{The voltage profile is linear during a short interval, i.e. $dE/dc$ is a constant}. The true value of the $dE/dx$ term is calculated from the open circuit voltage profile in \eqref{eq:ocv}.  Although the SOC interval during a current pulse is relatively small (1.785 \% SOC change in a GITT current pulse), the minor variation in the $dE/dx$ term cannot be ignored and its impact is directly reflected in the $D_s$ estimation. This will be demonstrated in the next part. One more observation in Fig.~\ref{fig:GITT_ICI_full_soc_dEdx} is that the GITT method can provide a better estimation of the $dE/dx$ profile. This is expected, since in the GITT method, the $dE/dx$ term is estimated between two equilibrium states and it is not affected by the current direction, taking into account that the open circuit voltage hysteresis is not included in the model. However, the ICI method obtains $dE/dx$ during a galvanostatic charge and discharge, and therefore the estimation is slightly shifted depending on the current direction. As can be noted in Fig.~\ref{fig:GITT_ICI_full_soc_D_s} and Fig.~\ref{fig:GITT_ICI_full_soc_dEdx}, in the beginning of the lithiation and delithiation processes in the ICI test, both the $dE/dx$ term and the estimated $D_s$ value have a larger mismatch from the true value, but the agreement improves with the time. Despite some minor shifts in the $dE/dx$ estimation, the $D_s$ value obtained from the ICI method has a very good agreement with those calculated using the GITT method.

\begin{figure}
     \centering
    \includegraphics{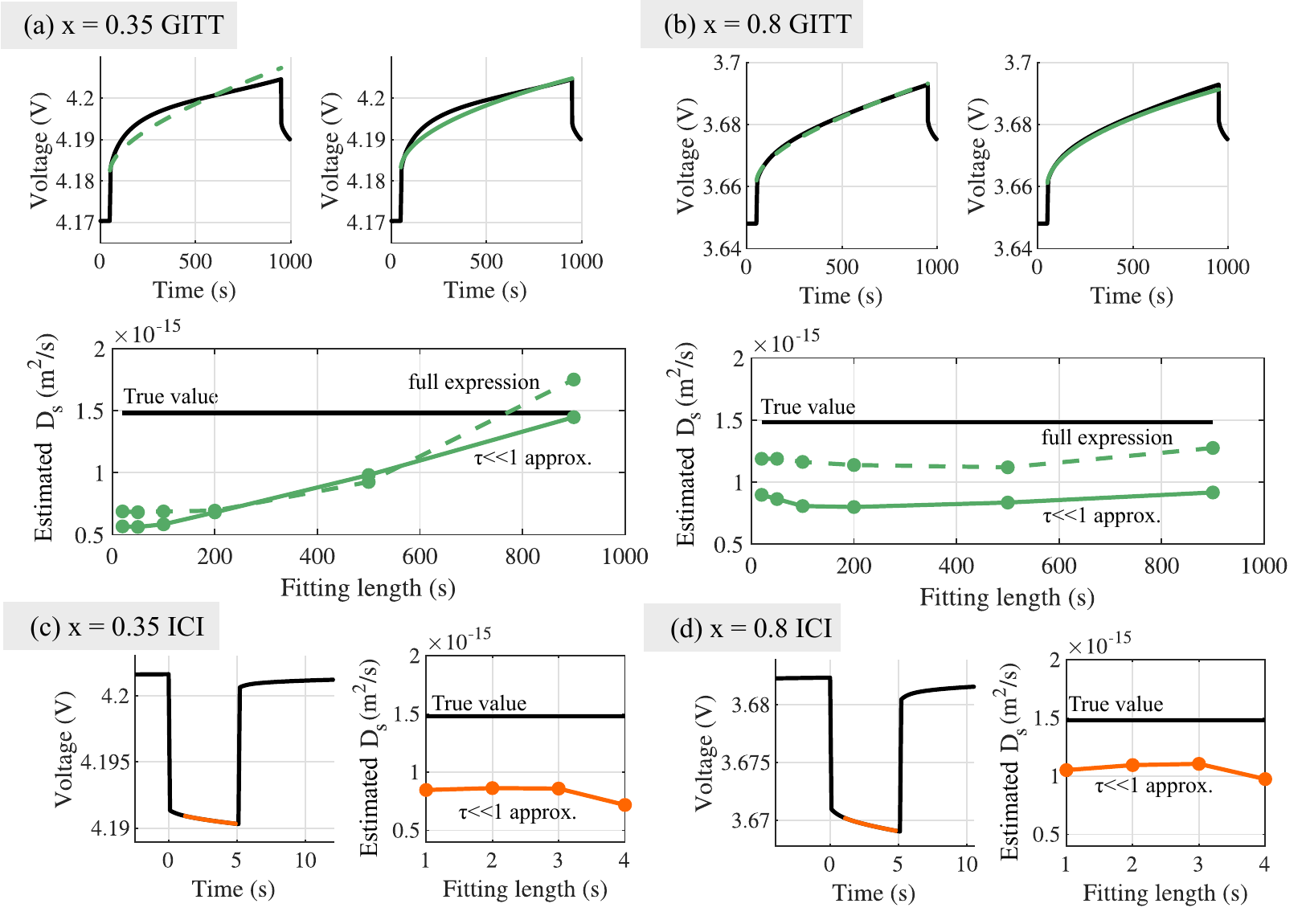}
    \caption{The voltage response of LiNi$_{0.8}$Mn$_{0.1}$Co$_{0.1}$O$_2$ and its fitting to obtain its diffusion coefficient $D_s$ in the GITT and ICI methods. The voltage response in black is the numerical solution simulated with a DFN model, and voltage response in colors are the analytical solutions with the fitted $D_s$, described in equations \eqref{eq:full_expression_fitting} as dashed line for the GITT method, \eqref{eq:tau<1 fitting} as the solid line for the GITT method and \eqref{eq:ICI_fitting} as the solid line for the ICI method, respectively.}
    \label{fig:dEdx_change_case1_case2}
\end{figure}

To demonstrate how the $dE/dx$ term affects the $D_s$ estimation, two cases at different SOC levels are studied in detail. In the first case ($x$ = 0.35), $dE/dx$ is changing rapid during the current pulse, while in the second case ($x$ = 0.8) $dE/dx$ is comparatively constant, which in turn fulfills the assumptions of the GITT derivation to a higher degree. Moreover, results with different fitting approaches and different fitting lengths are demonstrated as well. For the GITT method, the full expression fitting in \eqref{eq:full_expression_fitting} and with the $\tau \ll$ 1 approximation fitting in \eqref{eq:tau<1 fitting} are applied with a fitting length varying from 20 s to 900 s. When it comes to the ICI method, the expression in \eqref{eq:ICI_fitting} is used to find the diffusion coefficient, and the fitting length is from 1 s to 4 s. The voltage response in the first second is always excluded in the fitting to avoid the transient behaviours in other electrochemical processes in the system. The voltage response and its fitting are presented in Fig.~\ref{fig:dEdx_change_case1_case2} and the estimated diffusion coefficient are compared with the true value used in the DFN model.

It can be observed that the analytical solutions of Fick's law (colored lines) cannot capture the voltage response accurately when $x$ = 0.35 in Fig.~\ref{fig:dEdx_change_case1_case2}a, due to the relatively large variation of the $dE/dx$ term at this SOC level, as indicated in Fig.~\ref{fig:GITT_ICI_full_soc_dEdx}. As a consequence, the estimated $D_s$ when $x$ = 0.35 has a larger deviation from the true value of 1.48$\times$10$^{-15}$ m$^2$/s, for both the full expression fitting (dashed lines) and $\tau \ll$ 1 fitting (solid lines). It is seen in Fig.~\ref{fig:dEdx_change_case1_case2}a that this estimation can be improved with a longer fitting length. On the other hand, the analytical solutions (colored lines) highly resemble the numerical solution with a DFN model (black lines) when $x$ = 0.8 in Fig.~\ref{fig:dEdx_change_case1_case2}b, since the voltage profile is more linear during the current pulse, i.e. the $dE/dx$ term is close to a constant. Under these conditions, the estimated $D_s$ is closer to the true value, especially using the full expression fitting. The validity of the $\tau \ll$ 1 approximation was discussed in conjunction with Fig.~\ref{fig:Fickslaw_solution}, with the upper time axis calculated with the $r_s$ and $D_s$ values used in the model. The approximation is considered to be valid when $\tau <$ 0.0032, i.e. $t<$ 58.92 s. Therefore, the estimation discrepancy increases when the fitting length is increased from 20 s to 50 s and 100 s in Fig.~\ref{fig:dEdx_change_case1_case2}b, as this approximation is losing its validity. When the fitting length is increased even longer, the overpotential contributions from the charge transfer reaction and electrolyte mass transportation plays a role in the fitting, causing an opposite effect to the invalidity of the $\tau \ll$ 1 approximation, and improves the estimation in this case. Based on Fig.~\ref{fig:dEdx_change_case1_case2}a and Fig.~\ref{fig:dEdx_change_case1_case2}b, it can be concluded that the diffusion coefficient $D_s$ can be estimated more accurately when the voltage profile is more linear. The same conclusion can be made with the ICI simulation results in Fig.~\ref{fig:dEdx_change_case1_case2}c and Fig.~\ref{fig:dEdx_change_case1_case2}d, that the estimated $D_s$ with the ICI method is closer to the true value at $x$ = 0.8 with a constant $dE/dx$. Moreover, it can be observed that in the ICI method, it is preferable to keep the fitting length short to better fulfill the $\tau \ll$ 1 approximation. Despite the fact that the diffusion coefficient $D_s$ is slightly underestimated in most of the presented cases, this is considered as a fairly accurate estimation, with the maximum error only being 66\%.

\begin{figure}
     \centering
     \begin{subfigure}[b]{0.49\textwidth}
    \centering
    \includegraphics{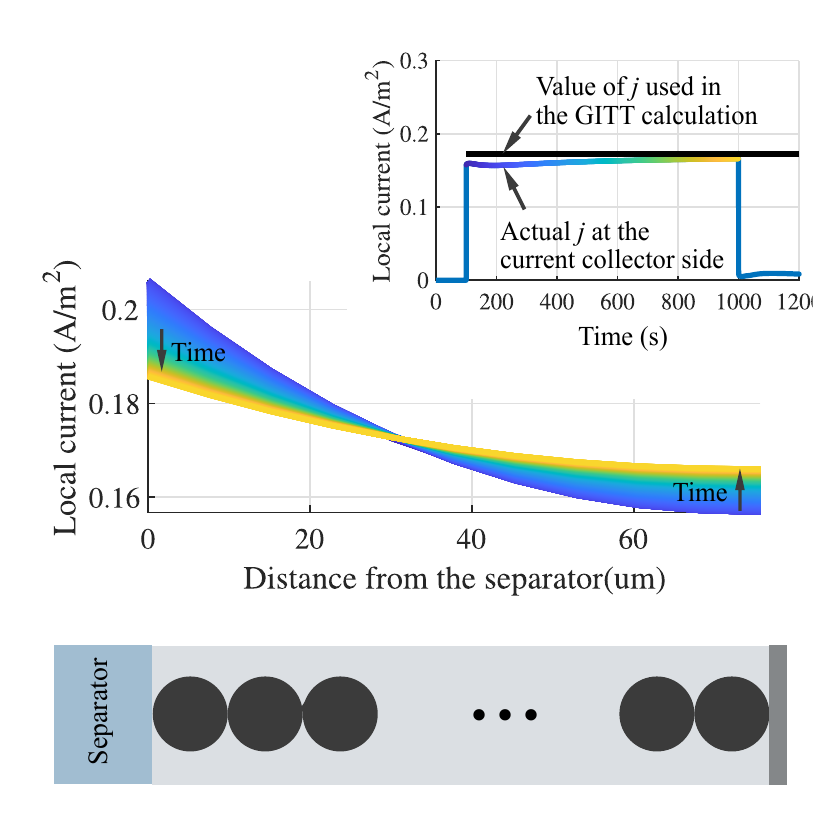}
    \caption{}
    \label{fig:Current_distribution}
     \end{subfigure}
     \hfill
     \begin{subfigure}[b]{0.49\textwidth}
    \centering
    \includegraphics{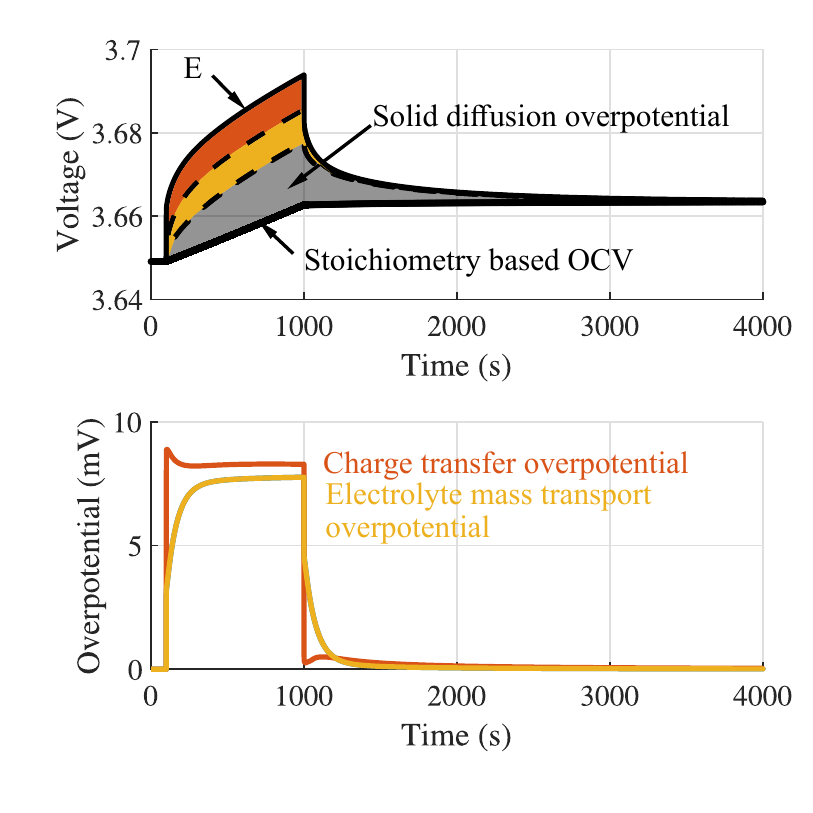}
    \caption{}
    \label{fig:Overpotential}
     \end{subfigure}
    \caption{During a current pulse at $x$ = 0.8, (a) the current distribution in the electrode during one GITT pulse and (b) Overpotential at the current collector side.}
\end{figure}

However, even when the voltage profile is linear at $x$ = 0.8, it is still not possible to obtain exactly the true value of $D_s$ irrespective of the fitting method and fitting length used. This can be explained with the violations of the other two assumptions in the derivation listed in the theory section: \textit{The current density at the particle surface is constant} and \textit{the over-potential caused by other electrochemical processes can be ignored}. To address these two issues, the current distribution in the electrode during one GITT pulse and the overpotential development are studied in detail in Fig.~6. In this analysis, the local current density and the overpotential at the particle closest to the current collector are studied, as it directly determines the measured terminal voltage. The current distribution during a current pulse in the GITT method at $x$ = 0.8 is shown in Fig.~\ref{fig:Current_distribution}. When the current is applied at $t$ = 0 s, the current density is higher at the separator side, since the electronic conductivity is much higher than the ionic conductivity and the current tends to flow via the least resistive path. With time passing, the current density becomes more even through the porous electrode. The actual local current $j$ at the current collector side is presented in Fig.~\ref{fig:Current_distribution}, and as can be seen, it is increasing with time. For the GITT fitting, the current is assumed to have an even distribution and the local current density $j$ is estimated to be $j = I/(S_a L)$ = 0.17 A/m$^2$ during a C/10 current pulse (plotted as the black line). It can be observed that in the beginning of the pulse, the actual current density at the current collector side is lower than 0.17 A/m$^2$, meaning that 0.17 A/m$^2$ is an overestimation of the local current density. This is one of the reasons for the disagreement between the estimated $D_s$ value and the true value. 

Besides the local current density, another factor causing this discrepancy is the overpotential caused by other electrochemical processes. As shown in Fig.~\ref{fig:Overpotential}, the measurable voltage response $E$ is the sum of the stoichiometry based voltage and the overpotential. In a DFN model, the overpotential is a consequence of three processes: the charge transfer reaction, electrolyte mass transport and the diffusion in the solid state. The charge transfer and solid diffusion overpotential are presented for the particle closest to the current collector, and the electrolyte mass transport overpotential is the difference between the electrolyte potential at the current collector point and electrolyte overpotential at the middle of the separator (the reference potential). Although it is only the solid diffusion overpotential that is the desired to use for the $D_s$ estimation, in reality it is not possible to separate it from the other two overpotentials. It can be observed in Fig.~\ref{fig:Overpotential} that both the charge transfer and electrolyte overpotential show a transient behaviour in the beginning of the pulse, which shall be avoided in the fitting. This is why the voltage response in the first second has been excluded in the previous demonstrations. These two overpotentials stabilize with time, and a higher electrolyte diffusion coefficient $D_l$ and a faster reaction rate will speed up the stabilization. In the later part of the current pulse, these two overpotentials constitute an offset in the voltage response while the solid diffusion overpotential is still developing with time. This offset will still impact the fitting result as it introduces errors in the $dE/d\sqrt{t}$ fitting.



\begin{figure}[!ht]
    \centering
    \includegraphics{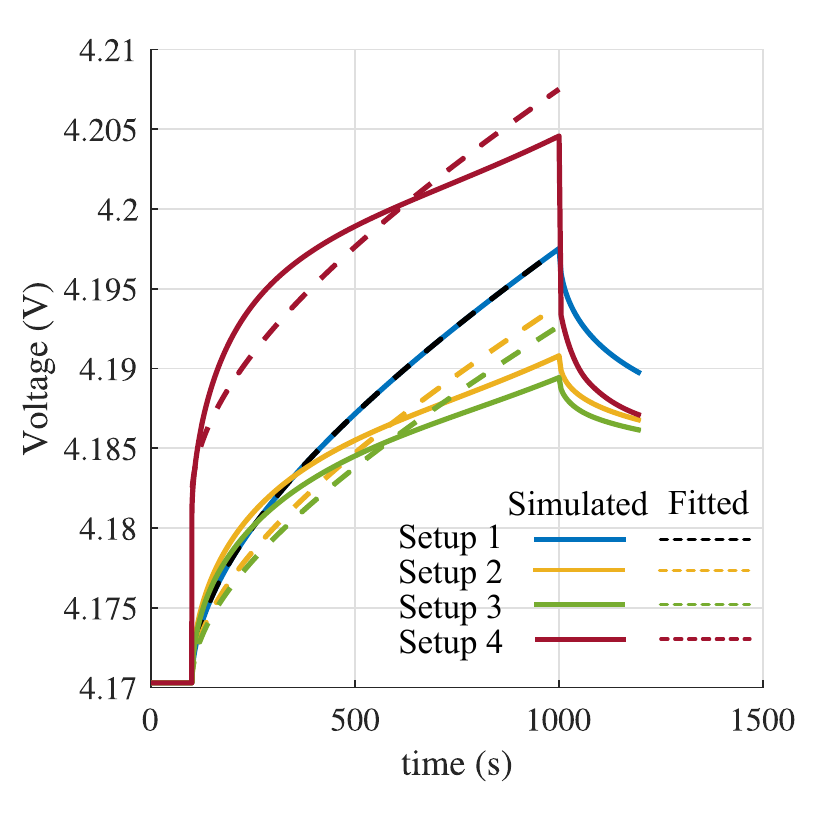}
    \caption{The voltage response during a current pulse simulated with a black-box configured according to Table.~\ref{tab:case_setup} (solid lines) and its fitting with analytical expressions with (10) (dashed lines).}
    \label{fig:Case_1_2_3_4}
\end{figure}

To further demonstrate how the violations of the last three assumptions in the theory section can affect the result, four setups are studied, from the most simplified scenario to a more realistic system. Four black-box models are configured according to Table~\ref{tab:case_setup} and the simulated voltage responses are plotted as solid lines in Fig.~\ref{fig:Case_1_2_3_4}. The simulated voltage response are then fitted with the analytical solutions, both with the full expression in (10), plotted as dashed lines, and the $\tau \ll 1$ approximation in (12), not shown in the figure. The first setup is an ideal scenario, where the voltage profile of the electrode material is perfectly linear, with a constant local current density and excluding the rest of the system. With this setup, the analytical solution of Fick's law can precisely represent the voltage response, as shown in Fig.~\ref{fig:Case_1_2_3_4}, and the true $D_s$ value can be accurately estimated without any deviations using the full expression fitting \eqref{eq:full_expression_fitting}. However, the $\tau \ll$ 1 fitting method leads to a 7.9 \% underestimation, since the approximation is conditionally valid. In the next setup, the voltage profile is a function of SOC, described in \eqref{eq:ocv} at $x$ = 0.35. The instant observation is that the analytical solution does not match the numerical solution anymore and the estimated $D_s$ values differ more than 40 \% compared to the true value. To complicate the system further, an uneven current distribution in the electrode is included in Setup 3. With this more realistic current distribution, the local current density at the current collector side is lower than the average value, as demonstrated in Fig.~\ref{fig:Current_distribution}, and the voltage response in Setup 3 is therefore slightly lower than what is simulated with Setup 2. In the final setup, the overpotential caused by other electrochemical processes are included in the terminal voltage. This setup has the same condition as in a DFN model and the simulated voltage profile is identical to what was presented in Fig.~\ref{fig:dEdx_change_case1_case2}a. Based on the fitting results in Table.~\ref{tab:case_setup}, it can be noted that more violations of these assumptions do not necessarily lead to a worse estimation. The exact impact depends on the combination of the parameters in the system, as well as the current magnitude, which will be demonstrated in the next part.

\begin{table}[!ht]
    \centering
    \caption{Four setups to investigate the impact of violations of assumptions.}
    \begin{tabular}{| c | c |c|c | c| c|} \hline
       & $dE/dx$ & Local current & \makecell{Overpotential caused  \\ by other processes} & \makecell{Fitted $D_s$ with \eqref{eq:tau<1 fitting} \\ between 1-20 s}  &  \makecell{Fitted $D_s$ with \eqref{eq:full_expression_fitting} \\ between 1-900 s}\\ \hline
    Setup 1  & Constant & Constant & Excluded &  1.36$\times$10$^{-15}$ m$^2$/s (-7.9\%) &  1.48$\times$10$^{-15}$ m$^2$/s (0\%) \\ \hline 
    Setup 2 & Varying & Constant & Excluded &  0.85$\times$10$^{-15}$ m$^2$/s (-42.6\%) &  2.15$\times$10$^{-15}$ m$^2$/s (+45.3\%) \\ \hline
    Setup 3 & Varying & Varying  & Excluded & 0.98$\times$10$^{-15}$ m$^2$/s (-33.6\%) &  2.53$\times$10$^{-15}$ m$^2$/s (+70.6\%) \\ \hline
    Setup 4 (DFN model) & Varying  & Varying  & Included &  0.57$\times$10$^{-15}$ m$^2$/s (-61.7\%) &  1.73$\times$10$^{-15}$ m$^2$/s (+16.6\%) \\ \hline

    \end{tabular}
    \label{tab:case_setup}
\end{table}

\begin{figure}[!ht]
     \centering
     \begin{subfigure}[b]{0.49\textwidth}
    \centering
\includegraphics{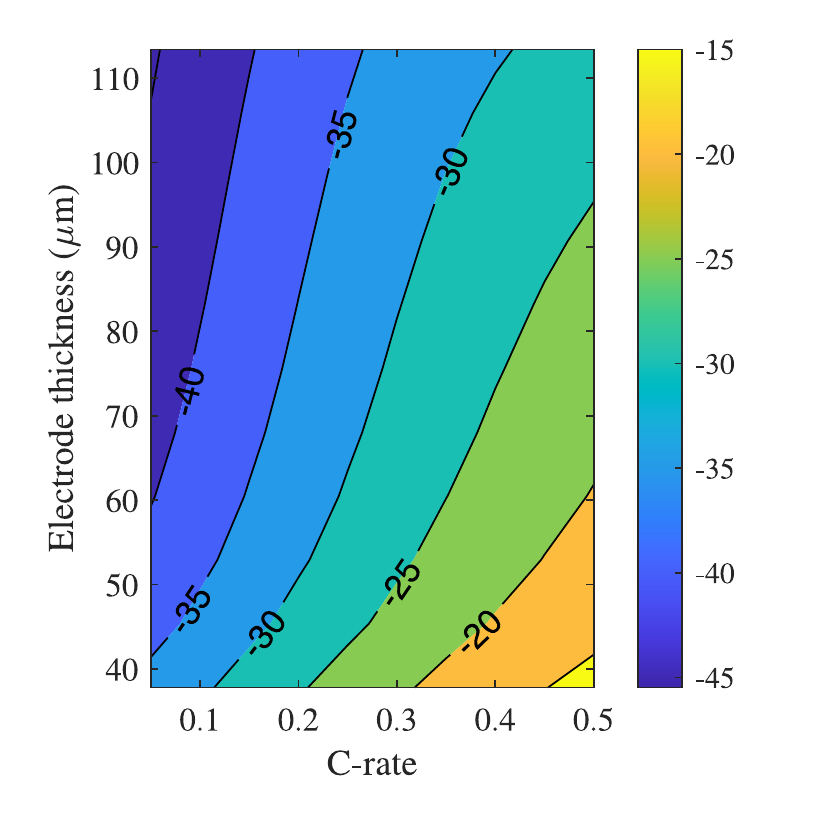}
    \caption{}
    \label{fig:D_s_sweep_current_thickness}
     \end{subfigure}
     \hfill
     \begin{subfigure}[b]{0.49\textwidth}
    \centering
    \includegraphics{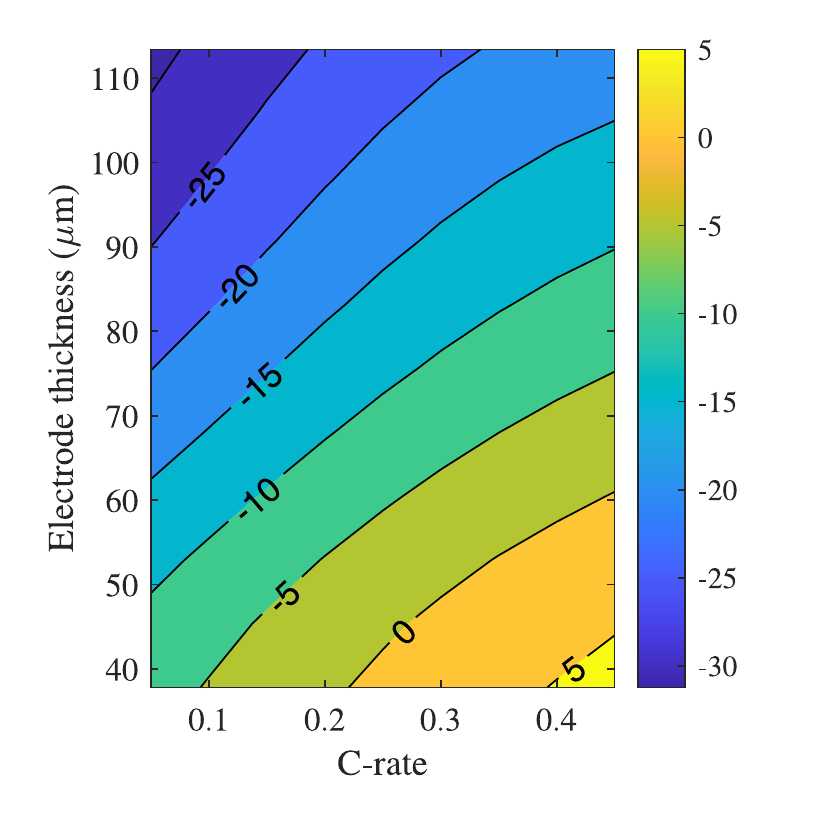}
    \caption{}
    \label{fig:D_s_sweep_current_thickness_higher_Dl}
     \end{subfigure}
    \caption{The difference between the estimated $D_s$ value and the true value with a variation in the electrode thickness and current magnitude with (a) the parameters listed in Table \ref{tab:parametersNMC811} (b) the diffusion coefficient in the electrolyte $D_l$ is increased by 10 times.}
\end{figure}

In the studies above, the estimation accuracy of $D_s$ is affected by the fitting approach and fitting length. Besides, the battery design, current magnitude and the rest of the system can have a large impact on the results as well. To illustrate such effects, the result of the estimated $D_s$ with a variation in the electrode thickness and current magnitude is presented in Fig.~\ref{fig:D_s_sweep_current_thickness}. The scenario investigated is that a current pulse is applied at $t$ = 0 s at $x$ = 0.8, and the voltage response between $t$ = 1 s and $t$ = 21 s is used for fitting with the $\tau \ll$ 1 approximation in \eqref{eq:tau<1 fitting}. The same study is then performed when the diffusion coefficient in the electrolyte $D_l$ in Table \ref{tab:parametersNMC811} is increased by 10 times; the results of this approach is shown in Fig.~\ref{fig:D_s_sweep_current_thickness_higher_Dl}. For both sets of parameters, the estimated value of $D_s$ is higher when employing a thinner electrode and a higher current magnitude. With a 40 $\mu$m thick electrode and C/10 current, the estimated $D_s$ is 32\% lower than the true value in Fig.~\ref{fig:D_s_sweep_current_thickness}, while the $D_s$ estimation becomes only 5\% lower than the true value in Fig.~\ref{fig:D_s_sweep_current_thickness}, meaning that an increased electrolyte diffusion coefficient $D_l$ can result in a higher estimation of the $D_s$ value. This does not necessarily mean that this can improve the estimation, since it can lead to an overestimation under some conditions. When the GITT method is used for estimating lithium-ion battery aging, a decreasing trend in the $D_s$ can be observed with the increase of the cycling numbers \cite{chien2021fast}. The result in Fig.~8 implies that apart from the active material degradation being the main reason, the decrease in $D_s$ could also be partly due to the degradation of the electrolyte and a decreased $D_l$. 


\section{Conclusions}

A theoretical validation of the GITT and ICI methods was demonstrated with a porous lithium-ion NMC811 electrode employing a DFN model. Although the theoretical derivation of the GITT and ICI methods are based on multiple assumptions which are most likely not entirely valid for this type of lithium-ion battery electrode, both methods can still provide a very good estimation of the $D_s$ value, with the maximum deviation being an underestimation of 66 \%. The estimation is more accurate when the voltage profile has a more linear shape, since the voltage response can then be better described with the analytical solution of Fick's law. For the GITT method, a longer fitting time can sometime provide a better estimation, while for the ICI method it is preferred to keep the fitting length short. With the parameter set used in this work, the solid-state diffusion coefficient is underestimated in most of the case. A higher current magnitude, a thinner electrode and a higher electrolyte diffusion coefficient can lead to a higher estimation for the $D_s$ value, but can also render an overestimation. For the purpose of identifying the solid-state diffusion coefficient, the ICI method has a significant advantage of being time-effective. It should be acknowledged that GITT can be used to study also other properties of an electrode material, including the open circuit voltage and hysteresis. However, the ICI technique is much more convenient and powerful and it is suitable to be incorporated as part of a reference performance test for aging diagnostics.


\section*{Acknowledgment}
The authors would like to thank Energimyndigheten (P42789-1) for the financing of this work. Y.-C.C. and DB acknowledges STandUP for Energy.


\nomenclature[A]{$T$}{Temperature, K}
\nomenclature[A]{$n$}{Number of electrons involved in the redox reaction}
\nomenclature[A]{$F$}{Faraday constant, 96485~s$\cdot$A/mol}
\nomenclature[A]{$i$}{Current density per electrode area, A/m$^2$}
\nomenclature[A]{$j_0$}{Exchange current density, A/m$^2$}
\nomenclature[A]{$\alpha$}{Transfer coefficient}
\nomenclature[A]{$\eta$}{Overpotential caused by the redox reaction, V}
\nomenclature[A]{$c$}{Lithium ion concentration, mol/m$^3$}
\nomenclature[A]{$c_{s,surf}$}{Lithium ion concentration on the particle surface, mol/m$^3$}
\nomenclature[A]{$k$}{Reaction rate constant}
\nomenclature[A]{$c_{s,max}$}{Maximum concentration in the intercalation material, mol/m$^3$}
\nomenclature[A]{$c_0$}{Initial concentration in the intercalation material, mol/m$^3$}
\nomenclature[A]{$t$}{Time, s}
\nomenclature[A]{$j$}{Charge transfer current density per surface area, A/m$^2$}
\nomenclature[A]{$\epsilon$}{Volume fraction}
\nomenclature[A]{$D$}{Diffusion coefficient, m$^2$/s}
\nomenclature[A]{$t_+^0$}{Transference number of lithium ion}
\nomenclature[A]{$\kappa$}{Electrolyte conductivity, S/m}
\nomenclature[A]{$L$}{Electrode thickness, m}
\nomenclature[A]{$r_s$}{Radius of solid particles, m}
\nomenclature[A]{$r$}{Radial distance in the particle, m}
\nomenclature[A]{$I$}{Given current density, A/m$^2$}
\nomenclature[A]{$E$}{Terminal voltage, V}
\nomenclature[A]{$f_A$}{Activity dependence}

\nomenclature[A]{$x$}{Stoichiometry, $x = c_s/c_{s,max}$}
\nomenclature[A]{$\lambda$}{Positive roots of $\lambda  = \tan (\lambda)$}
\nomenclature[B]{$s$}{Solid phase}
\nomenclature[B]{$l$}{Liquid phase}
\nomenclature[B]{$a$}{Anodic}
\nomenclature[B]{$c$}{Cathodic}
\nomenclature[B]{$0$}{Initial condition}


\printnomenclature

\ifCLASSOPTIONcaptionsoff
  \newpage
\fi



%

\bibliographystyle{IEEEtran}
\bibliography{reference.bib}

%








\end{document}